\documentclass{nlaauth}
\usepackage{graphicx}
\usepackage{amsmath}
\usepackage{amssymb}
\usepackage{url}

\begin{document}
\NLA{0}{0}{00}{00}{00}

\runningheads{V.\ E.\ Sacksteder} {$O{(N)}$ algorithms for
disordered systems}

\title{$O{(N)}$ algorithms for disordered
systems}

\author{V.~E.~Sacksteder\corrauth}

\address{
Vecchio Edificio Marconi,
Dipartimento di Fisica,\\
Universit\`a degli Studi di Roma "La Sapienza," \\
P. Aldo Moro 2, 00185 Roma, Italy }

\corraddr{vincent@sacksteder.com}




\noaccepted{}

\begin{abstract}
The past thirteen years have seen the development of many
algorithms for approximating matrix functions in $O{(N)}$ time,
where $N$ is the basis size.  These $O{(N)}$ algorithms rely on
assumptions about the spatial locality of the matrix function;
therefore their validity depends very much on the argument of the
matrix function. In this article I carefully examine the validity
of certain $O{(N)}$ algorithms when applied to hamiltonians of
disordered systems. I focus on the prototypical disordered system,
the Anderson model. I find that $O{(N)}$ algorithms for the
density matrix function can be used well below the Anderson
transition (i.e. in the metallic phase;) they fail only when the
coherence length becomes large. This paper also includes some
experimental results about the Anderson model's behavior across a
range of disorders.

\end{abstract}


\section{INTRODUCTION}
Certain matrix functions - the Green's function, the density
matrix, and the logarithm - are very important to science and
engineering. Where they are important, scientists are faced with a
computational bottleneck: evaluation of a matrix function
generally requires $O{(N^{3})}$ time, where $N$ is the basis size
and usually scales linearly or worse with the system volume.  This
prohibits computations of large systems, even with modern
computers. In particular, if a matrix function must be
recalculated at every step of a system's evolution in time, then
calculations with a basis size larger than $O{(1000 - 10000)}$ are
not practical.

In 1991 W. Yang introduced the Divide and Conquer algorithm, which
approximated a matrix function in $O{(N)}$ time\cite{Yang91}. This
stimulated the development of many other $O{(N)}$ algorithms
\cite{Ordejon98, Scuseria99, Goedecker99} which have met
considerable success, permitting for instance quantum dynamics
calculations of tens of thousands of atoms.  All $O{(N)}$
algorithms rely on special characteristics of the system under
study, and the question of their validity can be answered only
after having specified that system. To date, theoretical studies
of the validity of these algorithms have occurred almost
exclusively within the conceptual framework of ordered systems,
using ideas of metals, insulators, and band gaps\cite{Baer97,
Baer97b, Goedecker98, IsmailBeigi99, Goedecker99b, Goedecker99,
Zhang01, Koch01, He01, Taraskin02}. In this paper I carefully
examine the applicability of $O{(N)}$ algorithms to disordered
systems. I calculate a matrix function in two ways: with an
$O{(N)}$ algorithm, and via diagonalization. Comparison of the two
results provides some new insight into when disorder can make
$O{(N)}$ calculations feasible.

    This paper begins by introducing $O{(N)}$
    algorithms (section 2) and then explaining how to measure the
    errors which these algorithms induce (section 3.)
    Section 4 introduces disordered systems and the density matrix function,
    while section 5 gives theoretical estimates of $O{(N)}$ errors. I present my
    numerical results
    in section 6, and finish in section 7 with a short assessment of their reliability.

\section{O{(N)} Algorithms}
 All $O{(N)}$ algorithms make three basic assumptions:
 \begin{itemize}
\item Existence of a Preferred, Local Basis. It is assumed that
the system is best described in terms of a localized basis set. I
here define a basis as localized if for any basis element ${| \psi
\rangle}$, only a small number of positions $\vec{x}$ satisfy
${\langle \vec{x} | \psi \rangle} \neq 0$.

\item Existence of a Distance Metric.  There must be a way of
computing the physical distance between any two basis elements
${|\psi\rangle}$ and ${|\acute{\psi}\rangle}$.

 \item Locality of both the Matrix Function and its Argument.
 I call a matrix $A$ local if, for every pair
 of basis elements $| \vec{x}
 \rangle$ and $| \vec{y} \rangle$ that are far apart,
  ${{\langle \vec{x} |} A {| \vec{y} \rangle}}
= 0$.  Throughout this article I choose a simple criterion for
being far apart: comparison with a radius $R$.
\end{itemize}

 There are a
number of ways for an $O{(N)}$ algorithm to exploit the three
basic assumptions. In this article I focus on the class of
algorithms based on basis truncation. This class includes Yang's
Divide and Conquer algorithm\cite{Yang91}, the "Locally
Self-Consistent Multiple Scattering" algorithm\cite{Abrikosov96,
Abrikosov97}, and Goedecker's "Chebychev Fermi Operator
Expansion"\cite{Goedecker94, Goedecker95}, which I will henceforth
call the Goedecker algorithm. Basis truncation algorithms break
the matrix function into spatially separated pieces. Given the
position of a particular piece, the basis is truncated to include
only elements close to that position, and then the matrix function
is calculated within the truncated basis. Thus, for any generic
matrix function $f{(H)}$, a basis truncation algorithm calculates
$ {{\langle \vec{x} |} {f{(H)}} {| \vec{y} \rangle}} = {{\langle
\vec{x} |} {f{(P_{\vec{x}, \vec{y}} H P_{\vec{x}, \vec{y}})}} {|
\vec{y} \rangle}}$, where $P_{\vec{x}, \vec{y}}$ is a projection
operator truncating all basis elements far from $\vec{x}$,
$\vec{y}$. There may be also an additional step of interpolating
results obtained with different $P$'s, but I will ignore this. In
this paper I choose $P$ to be independent of the left index
$\vec{x}$, and truncate all basis elements outside a sphere of
radius $R$ centered at $\vec{y}$.

Basis truncation algorithms vary only in their choice of how to
evaluate the function $ {f{(P_{\vec{x}, \vec{y}} H P_{\vec{x},
\vec{y}})}} $. Specifically, Yang's algorithm calculates $f$ in
terms of the argument's eigenvectors, the Goedecker algorithm does
a Chebyshev expansion of $f$, and the "Locally Self-Consistent
Multiple Scattering" algorithm calculates the argument's resolvent
or Green's function and then obtains $f$ via complex integration.
Because these approaches are all mathematically equivalent when
applied to analytic functions, they should all converge to
identical results, as long as one makes identical choices of which
matrix function to evaluate, of how to break up the function, of
which projection operator to use, and of a possible interpolation
scheme. Moreover, given an identical choice of matrix function,
variations in the other choices should obtain results that are
qualitatively the same.  In this paper I use the Goedecker
algorithm, but I want to emphasize that the results obtained here
apply to the whole class of basis truncation algorithms.

The Goedecker algorithm is essentially a Chebyshev expansion of
the matrix function.  As long as all the eigenvalues of the
argument $H$ are between $1$ and $-1$, a matrix function may be
expanded in a series of Chebyshev polynomials of $H$: $f{(H)}
\cong { \sum_{s=0}^{S} {c_{s} T_{s}{(H)}}}$.  The coefficients
$c_{s}$ are independent of the basis size, and therefore can be
calculated numerically in the scalar case.  The Chebyshev
polynomials can be calculated in $O{(N)}$ time using the recursion
relation $T_{s+1} = {{({2 H T_{s}})}
 - T_{s-1}}$, $T_{1} = H$, $T_{0}$ = 1.  (Of course, one must also
 bound the highest and lowest eigenvalues of $H$ and then
 normalize.  In practice very simple heuristics are sufficient for
 estimating these bounds.)  If the matrix function $f{(H)}$ has a
 characteristic scale of variation $\alpha$, then the error induced by the Chebyshev
 expansion  is controlled by an exponential with argument of order
 $-\alpha S$.

\section{Measuring The Error} Previous efforts to test
numerically the accuracy of $O{(N)}$ algorithms have been confined
to evaluations of whether the overall physical predictions are
reasonable, and tests of convergence with respect to the
localization radius $R$ and any other parameters.  In contrast, in
this paper I compute a matrix function using both an $O{(N)}$
algorithm and an algorithm based on diagonalization, and then
compare the results.

This careful comparison required development of a metric for
comparing two matrices. First, note that the dot product used for
comparing two vectors can be easily generalized to matrices:
$${MDP{(A, B)}} \equiv {Tr{(AB)}} =
{\sum_{\vec{x},\vec{y}}{{\langle \vec{x} | A | \vec{y}
\rangle}{\langle \vec{y} | B | \vec{x} \rangle}}} $$ If $A = B$,
this matrix dot product is just square of the Frobenius norm, one
of the traditional norms for matrices. Note also that matrix dot
product is invariant under change of basis. Moreover, it is simple
to show that ${-1} \leq {\frac{MDP{(A, B)}}{\sqrt{{MDP{(A, A)}}
{MDP{(B, B)}}}}} \leq {1}$, so one may define the angle $\theta$
between two matrices as the arcsin of this quantity. Bowler and
Gillan\cite{Bowler99} justified this, showing that the concept of
perpendicular and parallel matrices is valid and useful.

However the matrix dot product is not quite suited to my needs.
$O{(N)}$ algorithms have a preferred, local basis, and thus are
not well matched by a basis invariant measure.  Moreover, the
matrix functions which they compute are expected to agree best
with the exact matrix functions close to the diagonal, and to
agree not at all outside the truncation radius. Therefore, a more
sensitive metric is needed, one that distinguishes different
distances from the diagonal.  I define the Partial Matrix Dot
Product as:

$${MDP{(A, B, \vec{x})}} \equiv {\sum_{\vec{y}}{{\langle \vec{y} |
A | {\vec{y} + \vec{x}} \rangle}{\langle \vec{y} | B | {\vec{y} +
\vec{x}} \rangle}}} $$

The argument $\vec{x}$ of this dot product allows me to obtain
information about agreement at displacement $\vec{x}$ from the
diagonal.  It is still valid to call this a dot product, because
the magnitude of $\frac{MDP{(A, B, \vec{x})}}{\sqrt{{MDP{(A,
A,\vec{x})}} {MDP{(B, B, \vec{x})}}}}$ is bounded by one, and thus
one can compute a displacement-dependent angle $\theta{(\vec{x})}$
and relative magnitude $ m{(\vec{x})}$. The partial matrix dot
product has a simple sum rule relating it to the full matrix dot
product: $MDP{(A,B)} = {\sum_{\vec{x}} {MDP{(A, B, \vec{x})}}}$.

In my results I actually compute another dot product, an angular
average $MDP{(A, B, r)}$ over all $\vec{x}$ satisfying ${r} =
{|\vec{x}|}$.

 \section{The Density Matrix}
 In this paper I restrict my attention to a single matrix
 function, the density matrix.  This function is very important in
 quantum calculations of
 electronic structure in atoms and molecules, where its argument
 is the
 system's Hamiltonian, its diagonal
 elements describe the charge density, and its off-diagonal
 elements are used to compute forces on the atoms.  Eigenvalues of
 the Hamiltonian give the energies of their corresponding
 electronic states, and I
 will use the words eigenvalue and energy interchangeably
 throughout the rest of this paper.

 The density matrix function $\rho{(\mu, T, H)}$
 is basically a projection operator
 which deletes eigenvectors having energy $E$ larger than the
 Fermi level $\mu$.
  Here I use the following form:
$$\rho {(\mu, T, H)} = {\frac{1}{2} {Erfc}{(\sqrt{2}{(\frac{H -
{\mu I}}{T})})}}$$
  For physical reasons, it is not quite a projection
  operator: it has a transition
 region around $\mu$ of width proportional to the temperature $T$
 where its eigenvalues interpolate between $0$ and $1$.
The error induced by a Chebyshev expansion is controlled by an
exponential with argument proportional to ${-T S} / { \Delta}$,
where $\Delta$ is the size of the energy band and $S$ is the
number of terms in the Chebyshev expansion\cite{Goedecker94}.

 The density matrix is well suited to $O{(N)}$ algorithms.  As $\mu$ becomes
large, it converges to the identity. Moreover, it is invariant
under unitary transformations acting on the set of eigenvectors
with energies below the Fermi level\cite{Scuseria99}. Even when
the Hamiltonian's eigenvectors are not a local basis set, often a
unitary transformation can be found which maps them to a basis
which is localized.  If such a transformation exists, the density
matrix is localized.

Several papers have examined density matrix locality in the
context of ordered systems; i.e. ones whose Hamiltonians possess
lattice translational invariance\cite{Baer97, Baer97b,
Goedecker98, IsmailBeigi99, Goedecker99b, Goedecker99, Zhang01,
Koch01, He01, Taraskin02}. (Lattice translational invariance can
be expressed quantitatively as $ {{\langle \vec{x}|} H { | \vec{y}
\rangle}} = {{\langle \vec{x} + \vec{\Delta}|} H { | \vec{y} +
\vec{\Delta} \rangle}}$ for all $\vec{\Delta}$ located on an
infinitely extended lattice.) It is well known that the
eigenvalues of such systems are arranged in bands separated by
energy gaps where there are no eigenvalues, and
 that the eigenvectors are extended through all
space.  Notwithstanding the nonlocality of the eigenvectors, there
are strong arguments for localization in all ordered systems. If
the system is metallic (meaning that the Fermi level lies in one
of the bands of eigenvalues) and the temperature is zero, then in
a three-dimensional system the density matrix is expected to fall
off asymptotically as ${R^{-2}}$, where $R$ is the spatial
distance from the diagonal. A non-zero temperature multiplies this
by an exponential decay.  If instead the system is an insulator,
then the density matrix should decay exponentially even at $T =
0$.

Most systems of physical interest do not exhibit lattice
translational symmetry.  In particular, many exhibit
inhomogeneities at scales much smaller than that of the system
itself.  These are termed disordered systems.   In this article I
study the prototypical disordered system, the Anderson
model\cite{Anderson58}. It describes a basis laid out on a cubic
lattice, one basis element per lattice site, and a very simple
symmetric Hamiltonian matrix $H$ composed of two parts:
\begin{itemize}
\item A regular part: ${\langle \vec{x}|H| \vec{y} \rangle} = 1$
if $\vec{x}$ and $\vec{y}$ are nearest neighbors on the lattice.
This term is, up to a constant, just the second order
discretization of the Laplacian; its spectrum consists of a single
band of energies between $-2D$ and $2D$, where $D$ is the spatial
dimensionality of the lattice.

\item A disordered part: Diagonal elements $<\vec{x}|H|\vec{x}>$
have random values chosen according to some probability
distribution.  In this article I choose to use the Gaussian
distribution
$${P{(V)}} = {\frac{1}{\sqrt{2 \pi \sigma}} {\exp{(\frac{- V^{2}}{2 \sigma}
)}}}$$ I call $\sigma$ the disorder strength; this is related to
the disorder strength used in the literature by a factor of
$\sqrt{12}$\cite{Slevin01, Bulka87}.
\end{itemize}

At small disorder strengths, the Anderson model is dominated by
its regular part; in particular the eigenvectors are extended
throughout the whole system volume. However, there is a small but
important departure from the ordered behavior: at the band edges
one finds a few eigenvectors with volumes much smaller than the
system volume. In fact, there is an energy $E_{LOC}$ such that any
eigenvector with eigenvalue $E$ satisfying ${|E|}
> E_{LOC}$ is localized.  On average these eigenvectors
 decay exponentially with the
spatial distance from their maximum\cite{Kantelhardt02}. As the
disorder strength is increased, $E_{LOC}$ gets smaller and
smaller; i.e. more and more of the energy band becomes localized.
At a critical disorder strength the whole energy band becomes
localized.  This phenomenon is called the Anderson transition; for
the Gaussian probability distribution used in this paper it occurs
at the critical disorder ${\sigma}_{c} = {{6.149} \pm
{0.006}}$\cite{Slevin01}.

Note that these statements must all be understood as regarding the
ensemble of Hamiltonians determined by the probabilistic
distribution of the disorder: for instance, I am stating that
above the critical disorder the subset of Hamiltonians with
unlocalized eigenvectors is vanishingly small compared to the
total ensemble size. Moreover, these statements are valid for an
infinite lattice; the mapping to computations on a finite lattice
is not always absolutely clear.

Studies of the locality of disordered systems have traditionally
concentrated on computations of the Green's function, not the
density matrix. It is expected that the average of the Green's
function should decay exponentially as $
\exp{(\frac{-R}{\tilde{R}})}$, where $\tilde{R}$ is called the
coherence length\cite{Economou84}. The density matrix, as we will
see, is closely related to the Green's function, so one may hope
that its average will also decay exponentially.  However, there
are two reasons why this hope may be unjustified: First, we do not
need to know whether the average of all density matrices is
localized, but instead whether each individual density matrix is
localized. The difference between the two could be significant.
Second, in a system below the critical disorder there will be
unlocalized eigenvectors, and one might therefore expect the
${R^{-2}}$ behavior typical of a metal.

Many $O{(N)}$ computations have treated systems which are
disordered\cite{Goedecker94, Zhang01, Schubert03}.  However, the
$O{(N)}$ literature contains little theoretical material about the
applicability of $O{(N)}$ algorithms to disordered systems. The
originators of the "Locally Self-Consistent Green's Function"
algorithm, which does not truncate the basis but instead does a
sort of averaging outside of a radius $r$, suggested that $r$
should be related to the coherence length\cite{Abrikosov96}, and
also to the error induced by their averaging\cite{Abrikosov97}.
Zhang and Drabold computed the density matrix of amorphous Silicon
using exact diagonalization and found an exponential
decay\cite{Zhang01}. In the next sections, I will first argue
theoretically and then show numerically that $O{(N)}$ basis
truncation algorithms are applicable to disordered systems,
including ones far below the Anderson transition.

\section{Estimating the Error}
The quantity of interest is the relative error,
\begin{equation} \label{Definee}
{e{(r, R)}} \equiv {\frac{MDP{(\Delta{(R)}, \Delta{(R)},
r)}}{MDP{(f, f, r})}}
\end{equation}
where $R$ is the radius of the truncation volume and $\Delta{(R)}$
is the difference between the exact matrix function $f$ and the
approximate matrix function $\tilde{f}{(R)}$.

 A first guess can be
made from the intuition that the absolute error
\begin{equation}\label{DefineE}{E{(r, R)}} = {MDP{(\Delta{(R)},
\Delta{(R)}, r)}} \end{equation}
 is probably
bounded by its value at the boundary of the truncation region.
This allows a rough estimate of the relative error: $ {e{(r, R)}}
\lessapprox {\frac{E{(R, R)}}{MDP{(\tilde{f}, \tilde{f}, r})}}$,
suggesting that it can be made arbitrarily small if the matrix
function is localized. The numerator, however, is left undefined.
Perhaps it is reasonable to assume that on the boundary the
absolute error is equal to the approximate matrix function,
giving:

\begin{equation}
\label{RoughEstimate1} {e{(r,R)}} \lessapprox
{\frac{MDP{(\tilde{f}, \tilde{f}, R)}}{MDP{(\tilde{f}, \tilde{f},
r)}}}
\end{equation}

The following paragraphs develop further insight into the absolute
error by resolving $\Delta{(R)}$ into a multiple sum over dot
products between the argument's eigenstates $| \psi \rangle$ and
position eigenstates $|\vec{x}\rangle$ . Knowledge of the
normalization and asymptotic behavior of these states shed some
light on the magnitude of the matrix elements of the error: $
{\langle \vec{x} |} \Delta{(R)} {| {\vec{x} + \vec{r}} \rangle }$.

Basis truncation algorithms separate the basis into two projection
operators, $P_{A}$ for the part inside the localization cutoff
$R$, and $P_{B}$ for the part outside $R$. $P_{A}$ and $P_{B}$ are
then used to divide the argument $H$ into two parts: a part $H_{0}
= {{P_{A} H P_{A}} + {P_{B} H P_{B}}}$ which leaves $A$ and $B$
disconnected, and a boundary term connecting $A$ and $B$, $H_{1} =
{{P_{A} H P_{B}} + {P_{B} H P_{A}}}$.   The final result of a
basis truncation algorithm is just $P_{A} {f{(H_{0})}} P_{A}$.
Therefore, the error induced by a basis truncation algorithm is
entirely due to the boundary term $H_{1}$. In other words, ${P_{A}
{\Delta{(R)}} P_{A}} = {P_{A}{( {f{(H_{0} + H_{1})} - f{(H_{0})} }
)}P_{A}}$.

For matrix functions which are analytic on a region of the complex
plane which contains the poles of $H$ and $H_{0}$, an exact
equation for this boundary effect can be easily derived from the
Dyson equation. First define the Green's functions ${G{(E)}} =
{{(E - H)}^{-1}}$, ${G_{0}{(E)}} = {{(E - H_{0})}^{-1}}$.  Then
note that the matrix function can be obtained from the Green's
function through contour integration over the complex energy $E$:
${f{(H)}} = {\frac {1} {2 \pi \imath} \oint {G{(E)}f{(E)}}}$,
where the complex integral must contain the poles of $H$.  Next,
apply the Dyson equation $G = {G_{0} + {G H_{1} G_{0}}}$ twice to
obtain:
$$G = {G_{0} + {G_{0} H_{1} G_{0}} + {G_{0} H_{1} G H_{1}
G_{0}}}$$  This gives an exact relation between the correct
Green's function $G$ of the untruncated argument $ {H} = {H_{0} +
H_{1}}$ and the Green's function $G_{0}$ of the truncated argument
$H_{0}$. In order to obtain a similar relation for the matrix
function $f$, one must make the poles in this expression explicit
and then do a complex integration.  Define $| a \rangle $ and $| b
\rangle $ as two eigenvectors of $H_{0}$ which are both located
inside of the localization region, the set of $ | c \rangle $ as
the complete set of eigenvectors of $H$, and $E_{a}$, $E_{b}$, and
$E_{c}$ as their respective energies. Then:

\begin{equation} \label{EqExpand}
{{\langle \vec{x} |} {\Delta{(R)}} {| {\vec{x} + \vec{r}}
\rangle}} =
 {\int_{-\infty}^{\infty}\int_{-\infty}^{\infty}\int_{-\infty}^{\infty}{{{dE_{a}}{dE_{b}}{dE_{c}}}
 {{\langle \vec{x} |a \rangle} {{\langle a|}H_{1}{|c \rangle}}
  {{\langle c |} H_{1} {| b \rangle}} {\langle b | {\vec{x} + \vec{r} \rangle}}
{g{(E_{a},E_{b},E_{c})}}}}}
\end{equation}
where

$$ g{(E_{a}, E_{b}, E_{c})} = {{{n_{A}{(E_{a})}}{n_{A}{(E_{b})}}{n{(E_{c})}}}  \oint{{dE}f{(E)}
\frac{1}{E - E_{a}} \frac{1}{E - E_{b}} \frac{1}{E - E_{c}}}} $$

$n{({E})}$ is the spectral density ${\sum_{c}{\delta{(E-E_{c})}}}$
and is often approximated as a continuous function.  Similarly,
$n_{A}{({E})}$ is the spectral density of the eigenstates of
$H_{0}$ which are located inside the truncation region. If the
matrix elements and matrix function are well-behaved, then this
integral is also well-behaved.  Consider the integral ${\gamma} =
{\int \int \int {{{dE_{a}}{dE_{b}}{dE_{c}}}g{(E_{a}, E_{b},
E_{c})}}}$.  When $f$ is the density matrix and one uses a simple
model with $n$ equal to a constant $\frac{N}{w}$ inside the energy
band $[{\frac{-\omega}{2}},{\frac{\omega}{2}}]$, this integral is
of order $\frac{N }{{\omega}^{2}} {(\frac{4 \pi}{3}R^{3})}^{2}$
when $\mu$ is inside the energy band and $0$ when it is outside
the band.

    Assuming that all eigenvectors of both $H$ and $H_{0}$
    are unlocalized, one can use Eq. \ref{EqExpand} to derive an
    upper bound on ${|{{\langle \vec{x} |} {\Delta{(R)}} {| {\vec{x} + \vec{r}}
\rangle}}|}$ of order $R^{4}$.  In the case of the density matrix
this is a gross overestimate.  Nonetheless Eq. \ref{EqExpand} can
teach three lessons:

\subsection{\label{LocalizedSystems} Localized systems} Suppose that all the
eigenvectors are bounded by $\exp{(-\frac{|{\vec{x} -
\vec{x_{0}}}|}{L})}$, where $\vec{x_{0}}$ is the origin of the
eigenvector and $L$ is the minimum decay length of the system.
Then one can use Eq. \ref{EqExpand} to prove that ${|{{\langle
\vec{x} |} {\Delta{(R)}} {| {\vec{x} + \vec{r}} \rangle}}|}$ is
bounded by a polynomial times $\exp{(-\frac{{2 R} - r}{L})}$ for
large $R$, that the absolute error $E{(r,R)}$ is bounded by a
polynomial times $\exp{(-\frac{{4 R} - {2 r}}{L})}$, and that the
relative error can be made arbitrarily small. This suggests that
in localized systems the absolute error $E{(r,R)}$ depends
exponentially on $r$. If so, then Eq. \ref{RoughEstimate1} is a
gross overestimate.

\subsection{\label{UnlocalizedSystems} Unlocalized systems}
If the eigenvectors are unlocalized, then the magnitudes of
${\langle \vec{x} |a \rangle}$ and ${\langle b | {\vec{x} +
\vec{r}} \rangle}$ have no strong dependence on $\vec{x}$ and
$\vec{r}$. This suggests that ${|{{\langle \vec{x} |}
{\Delta{(R)}} {| {\vec{x} + \vec{r}} \rangle}}|}$ is roughly
independent of the position, and that $E{(r, R)}$ is roughly
independent of $r$, thus providing partial justification of Eq.
\ref{RoughEstimate1}.

\subsection{\label{FiniteCoherenceLength} Finite coherence length}
Eq. \ref{EqExpand} suggests that in systems with a finite
coherence length $\eta$ the absolute error is reduced, via
reduction of the matrix elements ${{\langle a|}H_{1}{|c \rangle}}$
and ${{\langle c|}H_{1}{|b \rangle}}$.  I assume a very crude
model of the incoherence where the
 eigenvectors are broken into domains with constant phase, each
  domain of size $\frac{4 \pi}{3}{\eta}^{3}$.  The main effect
   is to decrease any integral over an
  eigenvector by a factor of $\sqrt{N_{\eta}}$, where $N_{\eta}$
  is the number of different domains where the integrand is
  non-zero.  $H_{1}$ touches about $\frac{4 \pi R^{2}} {4 \pi {\eta}^{2}}$
  such domains.  Therefore if $R > \eta$,
  ${{\langle a|}H_{1}{|c \rangle}} \propto {\frac{{\eta}}{R}}$ and
  ${E{(r,R)}} \propto {\frac{{\eta}^{4}}{R^{4}}}$.  Analytic
  calculations of the second moment of the matrix element confirm the same scaling law.

\section{Results}
I studied ensembles of Anderson hamiltonians at eleven disorder
values between ${\sigma} = {0.65}$ and ${\sigma} = {9.00}$,
including the critical disorder ${\sigma}_{c} = 6.15$.   In the
results presented here a truncation radius of $R = 5$ was used,
but the results are qualitatively similar to those obtained with
${R} = {1, 2, 3, 4}$.  A lattice size of ${16}^{3}$ was used, and
calculations with ${12}^{3}$ and ${20}^{3}$ lattices at the
critical disorder ${\sigma}_{c}$ indicate that finite size effects
are small. The largest such effect is an improvement of the basis
truncation algorithm's accuracy at smaller lattice volumes.  At
each disorder I calculated the density matrix at $13$ values of
the Fermi level $\mu$ ranging uniformly from $-12$ to $12$, which
covered the whole energy band at the lower disorders, and most of
it even at larger disorders.  Close to the edges of the energy
band the density matrix magnitude drops precipitously and the
other observables also change rapidly; the main results reported
here apply only to values of $\mu$ where the spectral density
remains high.

A low temperature ($T = {0.05}$) was chosen in order to minimize
any temperature effect.  A careful examination of the density
matrix's behavior at $\mu = 0$ showed that any temperature effect
was swamped by other effects. In particular, at low disorder the
density matrix's behavior is dominated by lattice effects. Because
of the low temperature a large number of Chebyshev terms was
needed; for each matrix I used a number of terms $S$ equal to $25$
times the total band width, which was enough to make the
truncation error quite small.

\begin{figure}
\includegraphics[width=5cm, height=8cm, angle=270]{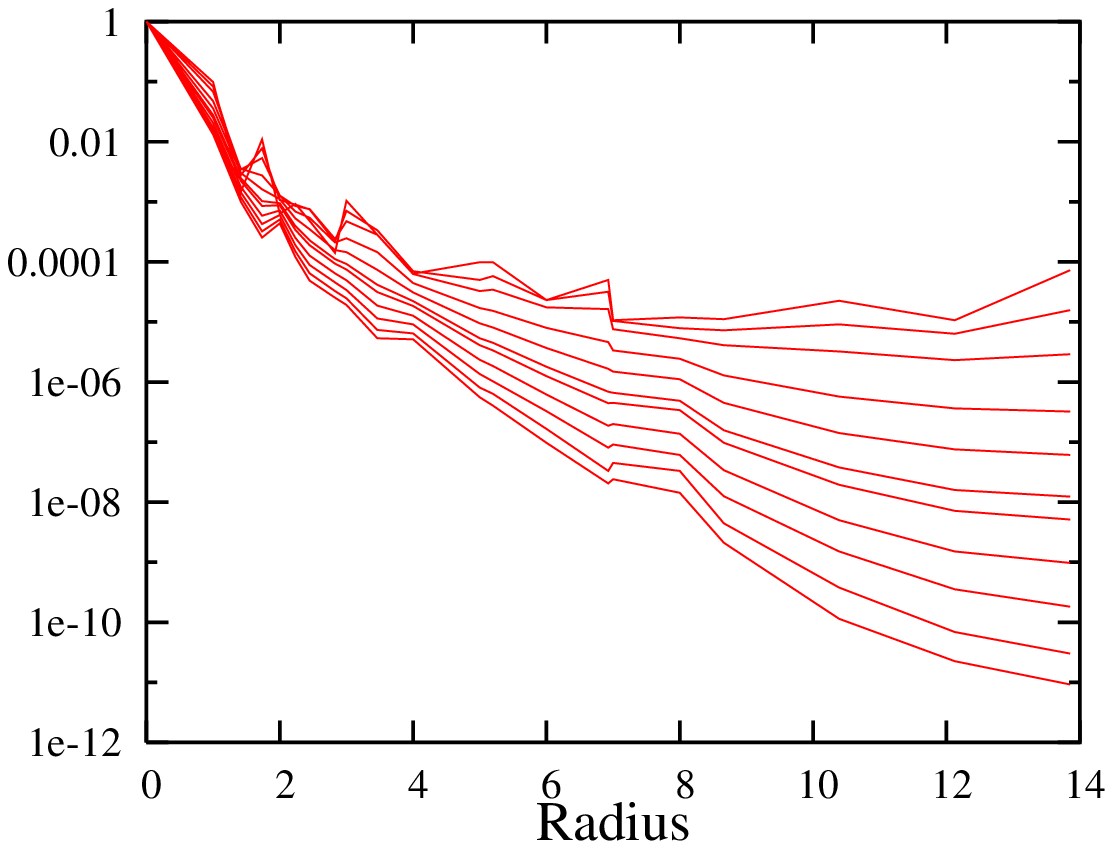}
\includegraphics[width=5cm, height=8cm, angle=270]{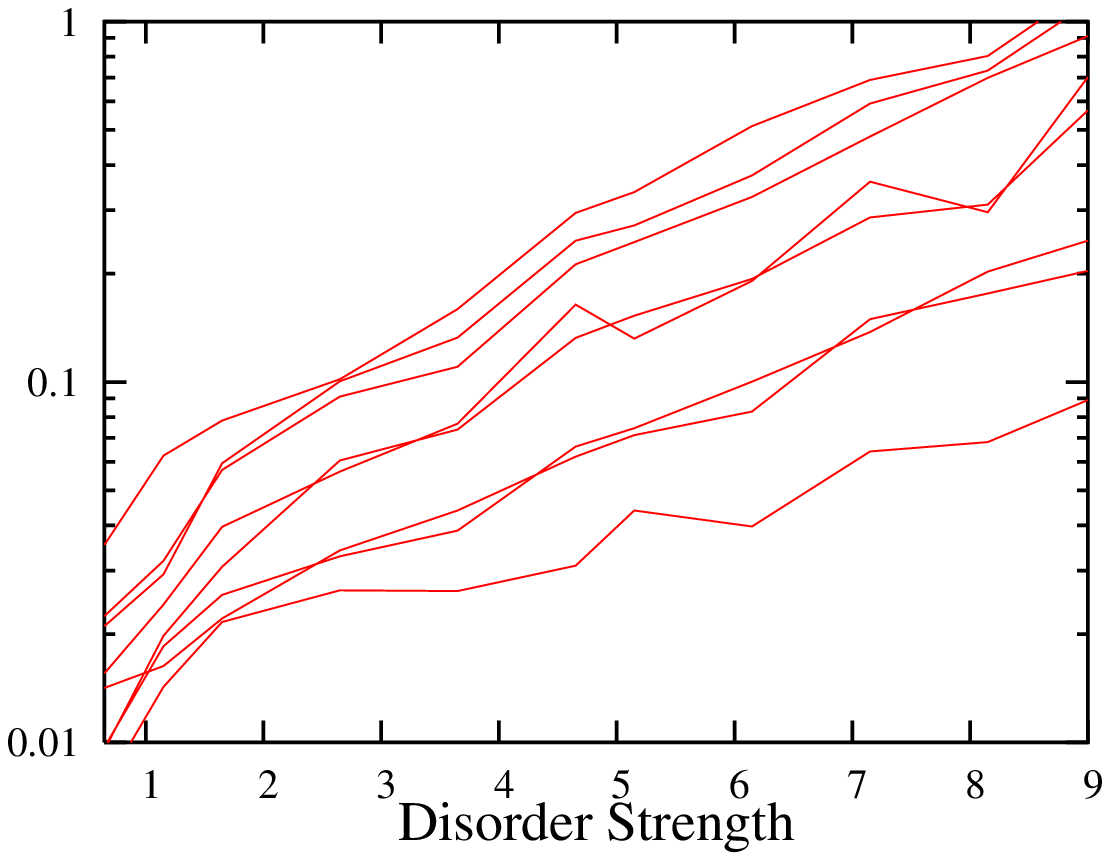}
\caption{\label{Figure2} The left graph shows the normalized
density matrix magnitude at ${\mu} = 0$.  Each line corresponds to
a different disorder strength between $0.65$ and $9.00$; lower
disorders are higher on the graph. The right graph shows the ratio
of the square root of the second moment to the mean at $\mu = 0$.
The different lines correspond to different radii $r = n\sqrt{3}$,
with smaller radii lower on the graph.}
\end{figure}

The left graph of figure \ref{Figure2} shows the normalized
density matrix magnitude $ {\frac{MDP{(\rho, \rho, r)}}{MDP{(\rho,
\rho, 0)}} }$ at ${\mu} = 0$. For $r > 0$ and $\sigma \geq 1.65$ a
good fit to this quantity can be obtained by $ {r^{-4}
\exp{(\frac{-2r}{\tilde{R}})}}$, where the coherence length is
given by $\tilde{R} = {{({0.057 \sigma} - {0.089} - {0.064
{\sigma}^{-1}})}^{-1}}$. Note the almost inverse relation between
the coherence length $\tilde{R}$ and the disorder strength
$\sigma$.  Lattice effects cause a systematic uncertainty in the
first constant $0.057$ of roughly $30\% $. Similar fits can be
obtained at other Fermi levels within the band ${|\mu|} \leq {6}$;
the first constant has a minimum at $\mu = 0$ and a total
variation of about $30\%$.

Now I consider the statistical distribution of the density matrix
magnitude. The right graph of figure \ref{Figure2} shows the ratio
of the square root of the second moment to the mean.  Note that
this ratio seems to grow roughly exponentially with the disorder
$\sigma$.  An examination of the kurtosis (the normalized fourth
moment of the statistical distribution) of the density matrix
magnitude shows that for $\sigma \geq {5.15}$ this quantity
becomes very large, starting at larger radii and larger Fermi
level $|\mu|$. These statistics suggest that at the Anderson
transition the statistical distribution of the density matrix
magnitude develops a long tail; it loses its self-averaging
property.

\begin{figure}
\includegraphics[width=5.5cm, height=8cm,angle=270]{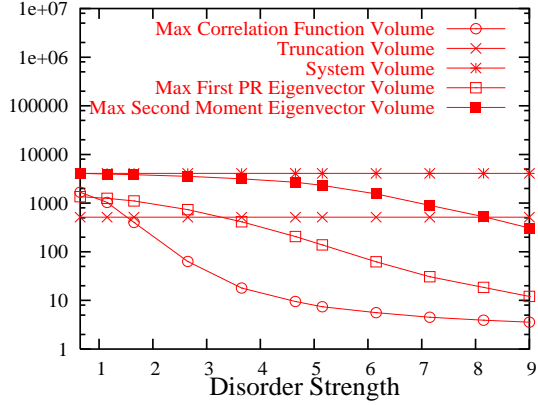}
\caption{\label{Figure3} Important volume scales.  Each volume
measure was averaged across the ensemble and across a small
interval of the energy spectrum.  Shown here are the maximum
values of these averages.}
\end{figure}

Figure \ref{Figure3} shows two measures of eigenvector volume. The
first is the inverse of the first participation ratio; i.e.
${({\sum_{\vec{x}}{{|{\langle \psi | \vec{x} \rangle}|}^2}})}^{2}
/ {\sum_{\vec{x}}{{|{\langle \psi | \vec{x} \rangle}|}^4}}$. This
quantity is a lattice friendly measure of volume because it has a
minimum value of one when  $ {\langle \psi | \vec{x} \rangle} $ is
a delta function and a maximum value of the system size when $
{\langle \psi | \vec{x} \rangle} $ is a constant. Figure
\ref{Figure3} shows that this volume becomes smaller than the
truncation volume in the range $\sigma = {3.00}$ to $\sigma =
{4.00}$.

My second volume measure is based on the second moment:
${({(2\pi)}^{D}{{det}{Q}})}^{\frac{-1}{2}} $ , where $Q$ is the
second moment (or quadrupole tensor) of ${{|{\langle \psi |
\vec{x} \rangle}|}^2}$.  Unlike the first volume measure, this
measure remains large even at large disorders, indicating that
each eigenfunction consists of several isolated peaks scattered
throughout the system volume.   This structure is caused by the
fact that states with similar energies will mix even if they are
connected by exponentially small matrix elements.  However, mixing
caused by such small matrix elements does not influence the
density matrix, because it essentially just induces a unitary
transformation of the mixed eigenvectors, and as we know the
density matrix is invariant under unitary transformations between
states that are all either less than $\mu$ or more than $\mu$.

 Figure \ref{Figure3} also shows the maximum value of the coherence volume;
 $V_{C} = {{max}_{E}{V{(E)}}}$, where $V{(E)}$ is the coherence
 volume.  I calculated this volume by first computing
the correlation function ${C{(\vec{x})}} =
{\int{{d\vec{k}}^{D}{{|{\langle \vec{k} | \psi \rangle
}|}^{2}{exp{(\imath {\vec{k} \cdot \vec{x}})}}}}} $, and then
applying my two volume measures to $ {C^{2}{(\vec{x})}}$.  Taking
the maximum value resolves an important ambiguity: at disorder
strengths $\sigma \leq {6.15}$ the coherence length shows two
peaks at energies ${|E|} = {6 - 8}$. These peaks become very
pronounced at $\sigma \leq 1.65$, where they are a factor of $10$
above the minimum.  Note that the coherence volume becomes small
much sooner than the eigenvector volume,
 in the range from $\sigma = {1.65}$ to $\sigma = {2.65}$.
For $\sigma \geq {1.15}$ and $V_{C} \gg 1$ it is roughly
proportional to ${{\sigma}^{-3}}$.  This fits well with the
density matrix's coherence length at large $\sigma$, but not at
small $\sigma$.

\begin{figure}
\includegraphics[width=5cm, height=8cm, angle=270]{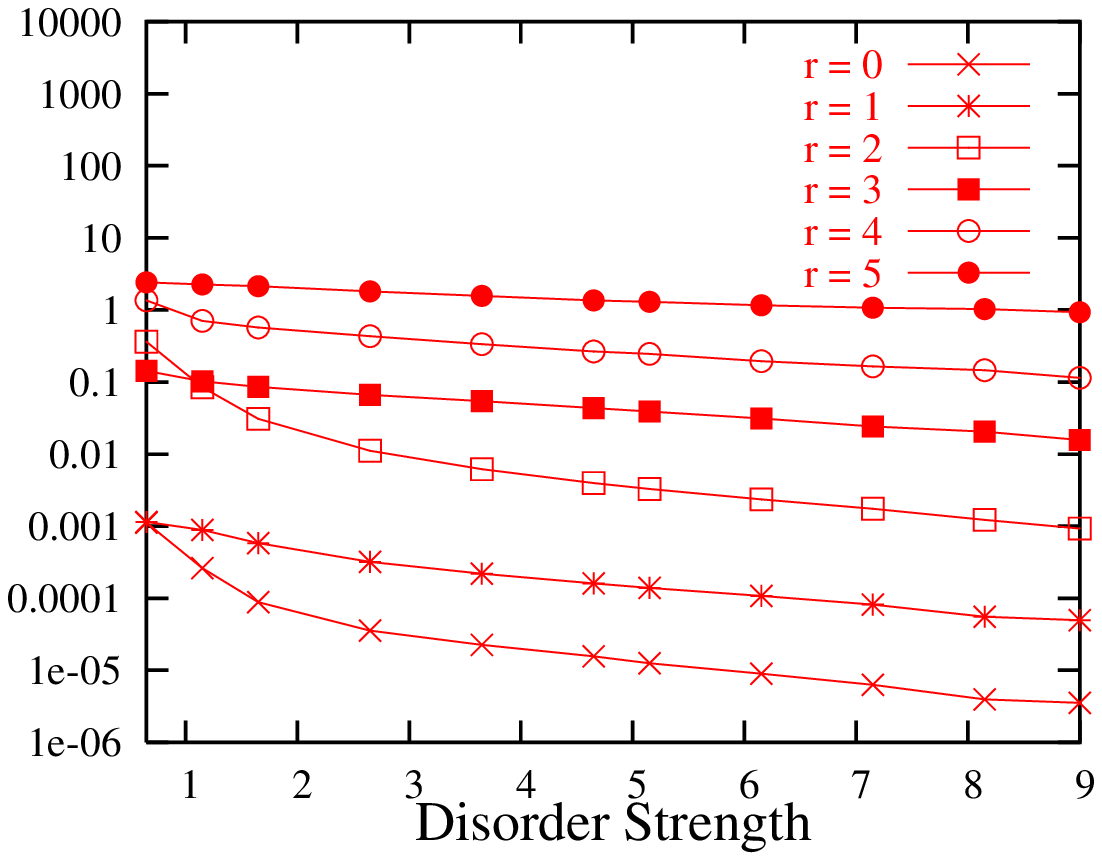}
\includegraphics[width=5cm, height=8cm, angle=270]{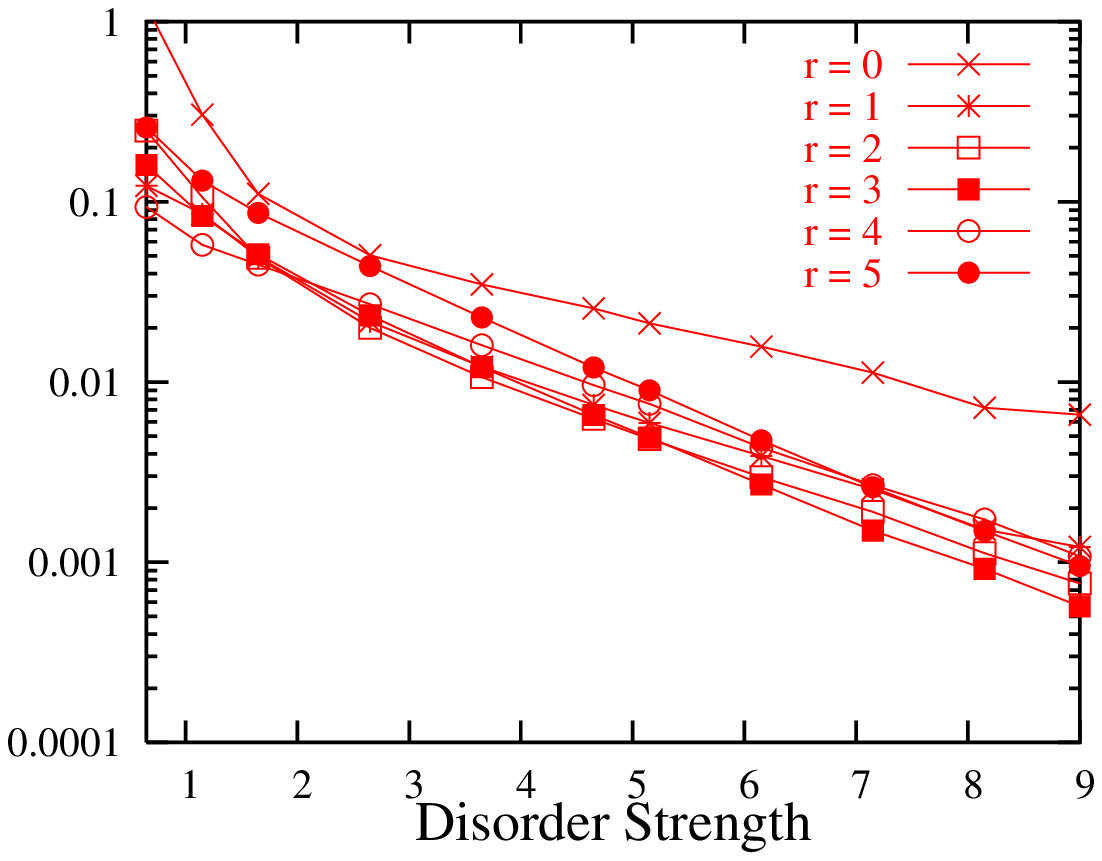}
\caption{\label{Figure4} Relative error $e{(r,R)}$ (eq.
\ref{Definee}) and absolute error error $E{(r,R)}$ (eq.
\ref{DefineE}) as a function of disorder. ${\mu} = {0}$.}
\end{figure}

Figure \ref{Figure4} shows the relative error and absolute error
as a function of disorder. The $O{(N)}$ algorithm begins to work
well at $r = {0,1}$ at quite small disorders, and the $r = 2$
relative error falls to $1\%$ at about $\sigma \thickapprox {3}$.
Clearly the $O{(N)}$ algorithm's success is controlled by the
coherence volume, not by the eigenvector volumes.

The overlapping lines in the right hand graph of figure
\ref{Figure4} confirm section \ref{UnlocalizedSystems}'s argument
that the absolute error $E{(r,R)}$ is roughly independent of $r$,
except at $r = 0$. The magnification at $r = 0$ is probably caused
by the density matrix's close relationship to the identity matrix.
It is more than compensated for by a corresponding magnification
of the density matrix at $r = 0$, so that the left hand graph
shows that the value of the relative error $e{(r,R)}$ at $r = 0$
is smaller than its value at $r=1$.

Section \ref{FiniteCoherenceLength} suggested that a small
coherence length may cause a decrease in the absolute error
$E{(r,R)}$ of order $\frac{{\eta}^{4}}{R^{4}}$.  The $R=5$ line in
the left hand graph of figure \ref{Figure4} shows that the
decrease is actually even more pronounced: at the boundary $r =
R$, $E{(r,R)}$ is of the same order of magnitude as
$MDP{(f,f,r)}$, which is controlled by an exponential.  This is
just the ansatz used to obtain Eq. \ref{RoughEstimate1}; my
results fully support the validity of Eq. \ref{RoughEstimate1}
except - as discussed before - at $r = 0$, and at the edges of the
energy band. Therefore accurate estimates of the error of an
$O{(N)}$ calculation may be obtained from the results of the
$O{(N)}$ calculation itself.

Section \ref{LocalizedSystems} showed that for large $R$ the
absolute error $E{(r,R)}$ is bounded by a polynomial times
$\exp{(-\frac{{4 R} - {2 r}}{L})}$, where $L$ is the decay length.
This suggests an $r$ dependence which is not confirmed by the
right hand graph of figure \ref{Figure4}, where it would cause a
splitting of the lines at $\sigma > {6.15}$. However, the previous
paragraph showed that $E{(r,R)}$ actually depends on $R$ as
$R^{-4} exp{(\frac{-2R}{\tilde{R}})}$, where $\tilde{R}$ is the
coherence length. Therefore as long as ${\tilde{R}} < {2 L}$, the
bound obtained in section \ref{LocalizedSystems} is automatically
verified at all $r$.  Moreover at small $R$ the unknown polynomial
in the bound may mask any $r$ dependence.

\section{Reliability of These Results}
I have already discussed the errors due to the finite lattice size
and to the truncation of the Cheybyshev expansion. While a precise
analytical and numerical treatment of both error sources is
possible, my checks indicate that they will have at most a
quantitative, not qualitative, effect on the results of this
paper. The main risks to the results probably lie in two areas:
finite ensemble size, and software reliability and
reproducibility.  I have taken steps to manage both issues:

\subsection{ Finite ensemble size.}  All results presented here
were obtained from ensembles of $33$ realizations, but repeating
the calculations with ensembles of $100$ realizations yielded the
same results. At the critical disorder the same quantities were
computed with three different lattice sizes ($100$ realizations at
both ${12}^{3}$ and ${16}^{3}$, and $10$ realizations at
${20}^{3}$), and the agreement is very good.  Graphing any
quantity across several disorders, one immediately notices that
there is little noise induced overlap of the two graphs.
Therefore, it seems likely that risks due to finite ensemble size
are under control.

\subsection{ Software Reliability and Reproducibility.}  I have tried
very hard to reduce this risk.  The software includes an automated
test suite which tests all computational functions except the
highest level output (graph printing) routines. Moreover, I have
taken pains to enable other researchers to easily reproduce and
check my results, simply by installing my software and the
libraries it depends on, compiling it with the GNU gcc
compiler\cite{GNUGCC}, and starting it running. The software, with
all needed configuration files, is available under the GNU Public
License\cite{GNUGPL}; check www.sacksteder.com for further
details.

\acks Special thanks to my advisor, Giorgio Parisi, for his
attentiveness, stimulation, and encouragement.  Thank you also to
Stephan Goedecker for sharing one of his $O{(N)}$ codes, to the
OXON group for sharing their code (though it remained unused), and
to the organizers of the 21st International Workshop on Numerical
Linear Algebra and its Applications for putting that event
together.


\end{document}